\documentclass{osa-article}

\journal{oe}

\articletype{Research Article}
\usepackage{caption}
\usepackage{lineno}
\linenumbers

\begin{document}

\nolinenumbers

\title{Multiple-access relay stations for long-haul fiber-optic radio frequency transfer}

\author{Qi Li, Liang Hu,\authormark{*} Jinbo Zhang, Jianping Chen, and Guiling Wu}

\address{State Key Laboratory of Advanced Optical Communication Systems and Networks, Department of Electronic Engineering, Shanghai Jiao Tong University, Shanghai 200240, China}

\email{\authormark{*}liang.hu@sjtu.edu.cn} 



\begin{abstract}
We report on the realization of a long-haul radio frequency (RF) transfer scheme by using multiple-access relay stations (MARSs). The proposed scheme with independent link noise compensation for each fiber sub-link effectively solves the limitation of compensation bandwidth for long-haul transfer. The MARS can have the capability to share the same modulated optical signal for the front and rear fiber sub-links, simplifying the configuration  at the repeater station and enabling the transfer system to have the multiple-access capability. At the same time, we for the first time theoretically model the effect of the MARS position on the fractional frequency instability of the fiber-optic RF transfer, demonstrating that the MARS position has little effect on system's performance when the ratio of the front and rear fiber sub-links is around $1:1$. We experimentally demonstrate a 1 GHz signal transfer by using one MARS connecting 260 and 280 km fiber links with the fractional frequency instabilities of less than $5.9\times10^{-14}$ at 1 s and $8.5\times10^{-17}$ at 10,000 s at the remote site and of $5.6\times10^{-14}$ and $6.6\times10^{-17}$ at the integration times of 1 s and 10,000 s at the MARS. The proposed scalable technique can arbitrarily add the same MARSs in the fiber link, which has great potential in realizing ultra-long-haul RF transfer. 
\end{abstract}

\section{Introduction}
Long-haul and ultra-stable radio frequency (RF) distribution is essential for numerous scientific and technical applications involving fundamental physics, radio astronomy and very long baseline interferometry (VLBI) \cite{1,2,3,4}. Owing to the influence of atmospheric environment, the traditional satellite-based time-frequency transfer methods such as Global Positioning System (GPS) or two-way satellite frequency transfer (TWSFT) are not satisfying the requirements of the modern high-precision applications \cite{5,6}. Compared with the satellite-based ones, the optical fiber is considered as an ideal medium for breaking the existing technological bottleneck and realizing high-precision and long-distance frequency transfer due to the advantage of low attenuation, wide bandwidth, immunity to electromagnetic interference, etc \cite{7}. In the past two decades, various excellent fiber-optic RF transfer schemes have been successively proposed and demonstrated \cite{8,9,10,Wang,12,13,14,sup1,sup2}.

To achieve ultra-stable fiber-optic RF transfer in quasi-national territories or even on a continental scale, the limitations of compensation bandwidth and signal-to-noise ratio (SNR) on the transfer system have to be resolved \cite{22}. Wang \textit{et al.} proposed a long-haul RF transfer scheme based on high-precision phase-locked loop (PLL) \cite{12}, demonstrating a 1007 km fiber-optic RF transfer in the single-span configuration by optimizing fiber link configurations such as the gain of erbium-doped fiber amplifiers (EDFAs), etc. Subsequently, Liu \textit{et al.} from the same group proposed a RF transfer scheme based on dual-PLLs, adopting PLL to improve the system’s SNR, which can effectively improve the short-term stability of the 1007 km fiber link \cite{13}. Although aforementioned single-span transfer schemes can obtain relatively low fractional frequency instability, the single-span transfer scheme deteriorates the noise suppression capability due to the decrease of the compensation bandwidth with the increase of the transfer distance. Furthermore, the optimization of the compensation system will become more difficult as the transfer distance increases \cite{22}. Thus, the cascaded transfer scheme with independent noise compensation for each fiber sub-link is more favourable for long-haul fiber-optic frequency transfer \cite{15}. Fujieda \textit{et al.} demonstrated the cascaded RF transfer scheme adopting two fiber sub-links over the total of the 204 km urban fiber link \cite{14}. Each fiber sub-link adopts an independent PLL-based active compensation technique, which can have the capability for stable operation. By adopting the similar technique, other research groups have also successively demonstrated a few cascaded RF transfer schemes for long-distance RF transfer applications \cite{14,15,16,17}. Unfortunately, the transfer stability degrades $\sqrt{N}$ times by cascading $N$ fiber sub-links \cite{16}. Additionally, more and more large-scale scientific and engineering facilities have an urgent demand for multiple-point RF access. However, the performance of existing multiple-point download schemes over the long-haul fiber link are mainly affected by the compensation bandwidth (depending on the length of the main link) and a large amount of attenuation introduced by the access nodes \cite{MA1,MA2, hu2021all, xue2021branching}. One intriguing question is how to design a multiple-access RF transfer scheme suitable for long-haul fiber link to improve the performance of the system in a cost-effective way.

In this article, we propose a long-haul fiber-optic RF transfer scheme by adopting novel multiple-access relay stations (MARSs). The proposed scheme with independent link noise compensation for each fiber sub-link mitigates the limitation of the compensation bandwidth for long-haul transfer. This novel MARS shares the same modulated optical signal for the front and rear fiber sub-links, not only effectively simplifying the configuration of RF transfer but also providing the multiple-access capability. Experimentally, we demonstrate the scheme by transferring a 1 GHz RF signal to the remote site (RS) via 260+280 km two fiber sub-links with the assistance of one MARS. In this configuration, the MARS can simultaneously compensate the front and rear fiber sub-links. Our experimental results demonstrate the stable RF signals can be obtained at the MARS and RS with the fractional frequency instability less than $5.6\times10^{-14}$ at 1 s and $6.6\times10^{-17}$ at 10,000 s, and $5.9\times10^{-14}$ at 1 s and $8.5\times10^{-17}$ at 10,000 s, respectively. To further verify this configuration, we theoretically analyze the effect of the MARS position on the system's performance. Promisingly, the proposed scalable technique can arbitrarily add the same MARSs in the fiber link for ultra-long-haul RF transfer.

\section{Principle}
\begin{figure}[htbp]
\captionsetup{labelfont=bf}
\centering\includegraphics[width=13cm]{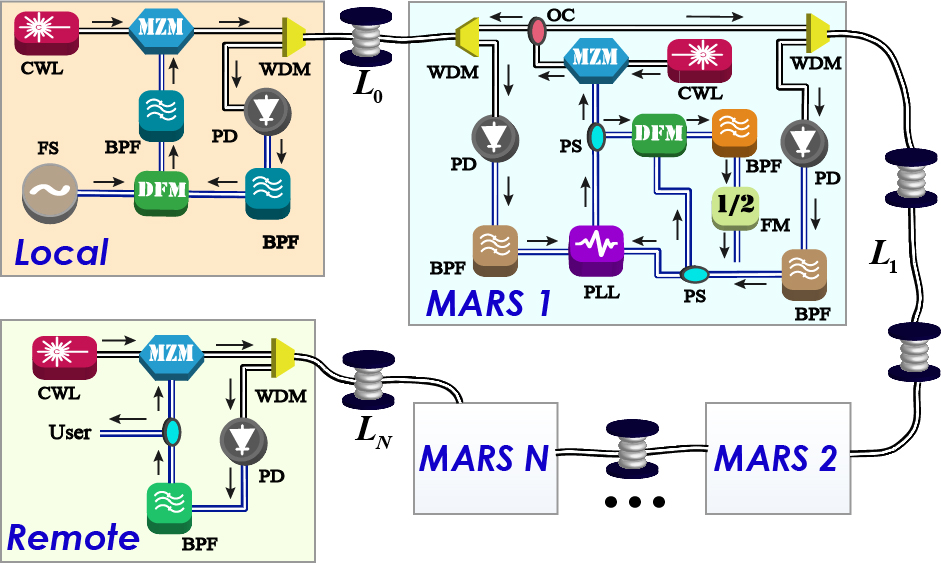}
\caption{Schematic diagram of long-haul optical fiber RF transfer with the assistance of $N$ MARSs. FS: frequency standard, CWL: continuous wave laser, PD: photodetector, MZM: Mach-Zehnder modulator, WDM: wavelength division multiplexer, Bi-EDFA: bidirectional erbium-doped fiber amplifier, OC: optical coupler, PLL: phase-locked loop, DFM: dual frequency mixer, BPF: band-pass filter, PS: power splitter, MARS: multiple-access relay station.}
\end{figure}

Figure 1 illustrates a schematic diagram of our proposed scheme for long-haul fiber-optic RF transfer by using $N$ MARSs. The local site (LS), $N$ MARSs and RS are connected by the $N+1$ fiber sub-links. Each MARS simultaneously transmits probe signal to the front and rear fiber sub-links to obtain the corresponding fiber-induced noise. At the LS, the frequency standard to be transmitted can be denoted as a cosine function without considering its amplitude, $E_r\propto cos(\omega_{r}t+\varphi_r$), where $\omega_r$ and $\varphi_r$ are its angular frequency and initial phase, respectively. 

At the any MARS, the probe signal produced by a tunable voltage-controlled oscillator (VCO) in the PLL unit consisting of mixer, PID controller and VCO, which can be denoted as,
\begin{equation}
 {E_1} \propto \cos ({\omega _{ck}}t + {\varphi _{ck}})\label{1},
\end{equation}
where $\omega_c$ and $\varphi_c$ stand for the angular frequency and initial phase of the VCO, the subscript number $k$ is an arbitrary number, indicating the corresponding the $k$th MARS. The frequency standard signal to be distributed and the probe signal of each MARS satisfy the frequency relationship of $\omega_r \approx{2\omega_{ck}}$. 

Although all MARSs have the same structure in our proposed scheme, the MARSs in different locations implement different functions, especially the first and last MARSs. At the first MARS, the probe signal is split into two parts. One part is modulated onto the continuous wave laser (CWL) by a Mach-Zehnder intensity modulator (MZM) biased at the quadrature point. The optically carried RF signal is divided into two parts, which are respectively transmitted to the LS and second MARS via the front and rear fiber sub-links. At the LS, the probe signal recovered by the photodetector (PD) can be written as, 
\begin{equation}
 {E_2} \propto \cos ({\omega _{\rm{c1}}}t + {\varphi _{p0}} + {\varphi _{c1}})\label{2},
\end{equation}
where $\varphi_{p0}$ represents the phase noise induced by the first fiber sub-link. By mixing $E_r$ with $E_2$ via dual frequency mixer (DFM) \cite{10}, the lower side-band signal is obtained and can be given by,
\begin{equation}
 {E_3} \propto \cos \left[ {\left( {{\omega _r} - {\omega _{{c1}}}} \right)t + {\varphi _r} - {\varphi _{p0}} - {\varphi _{c1}}} \right]\label{3}.
\end{equation}

At the LS, the $E_3$ is modulated onto an optical carrier by another MZM biased at the $V_{\pi/2}$ point and transmitted to the first MARS via the first fiber sub-link. Due to the high reciprocity of the forward and backward signal path \cite{Wang}, it is assumed that the phase noise introduced by the forward transmission signal via the same fiber link is the same as that of the backward transmission signal. Thus, the forward transfer signal detected by the PD at the first MARS can be depicted as, 
\begin{equation}
{E_4} \propto \cos \left[ {\left( {{\omega _r} - {\omega _{{c1}}}} \right)t + {\varphi _r} - {\varphi _{c1}}} \right]\label{4}.
\end{equation}

From the Eq. \ref{4}, it can be seen that the phase noise introduced by the first fiber sub-link is eliminated by passive compensation technique. At the first MARS, the backward probe signal from the second MARS recovered by the another PD can be denoted as,
\begin{equation}
{E_5} \propto \cos ({\omega _{{c2}}}t + {\varphi _{p1}} + {\varphi _{c2}})\label{5}.
\end{equation}

 The $E_4$ and $E_5$ signals are fed into the PLL unit of the first MARS. In the PLL unit, the DC error signal carrying the fiber-induced noise is obtained by mixing $E_4$ and $E_5$, resulting in,
 \begin{equation}
{V_{e1}} \propto \cos \left[ {\left( {{\omega _r} - {\omega _{c1}}- {\omega _{c2}}} \right)t + {\varphi _r} - {\varphi _{p1}} - {\varphi _{c1}}- {\varphi _{c2}}} \right]\label{6}.
\end{equation}

When the PLL unit is in-lock, the steady-state error is canceled by tuning VCO control voltage, i.e., $V_{e1}\rightarrow{0}$. Equation \ref {6} can be further written as,
\begin{equation}
{\omega _r} = {\omega _{c1}}+{\omega _{c2}},{\rm{ }}\quad{\varphi _{c1}} ={\varphi _{r}-{\varphi _{c2}-{\varphi _{p1}}}}\label{7}.
\end{equation}

By mixing $E_5$ and another part of probe signal with expression of $\omega_{c1}t+\varphi_{c1}$ via the DFM and taking the upper side-band, it can be seen that a stable RF signal with $\omega_{r}t+\varphi_r$ is obtained at the first MARS.

For the second MARS, the forward probe signal with $\omega_{c1}t+\varphi_{c1}+\varphi_{p1}$ and the backward probe signal with $\omega_{c3}t+\varphi_{c3}+\varphi_{p2}$ are sent into the PLL unit. When the second compensation device is in-lock, the following relationship can be obtained,
\begin{equation}
{\omega _{c1}}={\omega _{c3}},{\rm{ }}\quad{\varphi _{c1}+{\varphi _{p1}-{\varphi _{p2}}-{\varphi _{c3}}}}=0\label{8}.
\end{equation}

By substituting Eq. \ref{7} into Eq. \ref{8}, the Eq. \ref{8} can be rewritten as,
\begin{equation}
{\omega _r} = {\omega _{c2}}+{\omega _{c3}},{\rm{ }}\quad{\varphi _{c2}} ={\varphi _{r}-{\varphi _{c3}-{\varphi _{p2}}}}\label{9}.
\end{equation}

In a similar way, an expression of $ {\omega _r} = {\omega _{ck}}+{\omega _{ck+1}},{\rm{ }}\ {\varphi _{ck}} ={\varphi _{r}-{\varphi _{ck+1}-{\varphi _{pk}}}}$ similar with Eq. \ref{7} and Eq. \ref{9} can be obtained at the arbitrary $k$th MARS through $k-1$ iterations, except for the last MARS. By performing a similar mixing operation via the DFM, a stable RF signal can be obtained at arbitrary MARS. When the PLL unit at the last MARS is activating, we can obtain the relationship of,
\begin{equation}
{\omega _r} = 2{\omega _{cN}},\quad{\rm{ }}{\varphi _{cN}} = \frac{\varphi _r}{2}- {\varphi _{pN}}\label{10}.
\end{equation}

Noted that Eq. \ref{10} is the result of multiple iterations of the expression obtained for the previous $N-1$ MARSs. When $N$ MARSs are operating normally, the VCO signal in each MARS has the following relationship of $\omega_r=2\omega_{c1}=2\omega_{c2} \cdots=2\omega_{cN}$. At the RS, by substituting Eq. \ref{10} into the recovered RF signal with $\omega_{cN} t+\varphi_{cN}+\varphi_{pN}$, we can obviously find that a stable RF signal can be recovered at the RS.

\section{Experimental setup and results}
\subsection{Experimental setup}
We have demonstrated the proposed scheme with one MARS by adopting the experimental configuration as illustrated in Fig. 1. At the LS, the frequency standard (Rigol Inc., DSG 821) with $\omega_r$ is set to 2 GHz. The LS and MARS are connected by the front fiber sub-link with 260 km G.652 single-mode fibers (SMFs), denoted as $L_0$. The probe signal with $\omega_{c1}$ is set about 1 GHz at the MARS. The second fiber sub-link connecting the MARS and RS consists of 280 km SMFs (G.652) denoted as $L_1$. To suppress the impact of the chromatic dispersion, the proper dispersion compensating fibers (DCFs) are placed into the fiber-optic link. Table 1 elaborately indicates the length distribution of the SMF spools, the configuration of the DCFs and the attenuation of each section of the fiber link. It is noticeable that the third column in Table 1 stands for the nominal dispersion compensation length for G.652 SMF with the total length of the DCFs being approximately 54 km. We set the wavelengths of CWLs employed at the LS, MARS and RS to 1547.72nm, 1546.92 nm and 1547.72 nm, respectively. This scheme occupies only two International Telecommunication Standardization (ITU) channels (C37, C38) to get rid of the backscattering noise with the assistance of wavelength division multiplexers (WDMs). Each transmitted RF signal is modulated on the corresponding optical carrier by a LiNbO3 MZM (iXblue Inc., MXAN-LN-10) biased at the quadrature point and detected by the PD with 2 GHz bandwidth at each site.  A few home-made bidirectional erbium-doped fiber amplifiers (Bi-EDFAs) are placed into the fiber-optic link, in order to boost the fading bidirectional optical signals. The gain configuration of each Bi-EDFA is less than 25 dB to relieve the influence of the amplified spontaneous emission (ASE) noise and stimulated Brillouin scattering. The structure of the Bi-EDFA is similar with Ref \cite{18}, as shown in Fig. 2(a), which employs one unidirectional EDFA to complete the amplification of the bidirectional optical signals. We adopt seven home-made Bi-EDFAs in the transmission link, not only greatly simplifying the number of unidirectional EDFAs but also improving the symmetry of the single-fiber bidirectional transfer link in comparison with the Ref \cite{12}. Moreover, we select a PI controller (Newport Inc., LB1005) as the loop filter in the PLL unit. To suppress phase noise caused by the nonlinear-effect and RF leakage during the frequency mixing stage, the DFMs are adopted at the LS and MARS as illustrated in Fig. 2(b). In order to evaluate the performance of the proposed scheme, the frequency standard is divided by 2 at the LS and the duplicated 1 GHz signals at the RS (MARS) are converted to 10 MHz signals by a dual-mixer time difference method \cite{19}. The output signals are imported into the phase noise measurement instrument (Symmetricom Inc., TSC5120A) for the fractional frequency instability evaluation.

\begin{table}[htbp]
\centering
\begin{scriptsize}
\captionsetup{labelfont=bf}
\caption{\label{tab1}\textbf{Specific configuration of the optical fiber link}}
\begin{tabular}{cccc} \hline
Number & Length (km) & DCF (km) & Attenuation (dB)\\ \hline
1 & 80.8 & 80 & 22.1 \\
2 & 78.2 & 80 & 19.9 \\
3 & 50.1 & 40 & 12.7 \\
4 & 50.1 & 60 & 14.9 \\
5 & 50.4 & 60 & 13.4 \\
6 & 51 & 40 & 14.1 \\
7 & 80 & 80 & 20.6 \\
8 & 49.2 & 60 & 14.1 \\
9 & 50.1 & 40 & 12.9 \\\hline
\end{tabular}
\end{scriptsize}
\end{table}

\begin{figure}[htbp]
\centering\includegraphics[width=12cm]{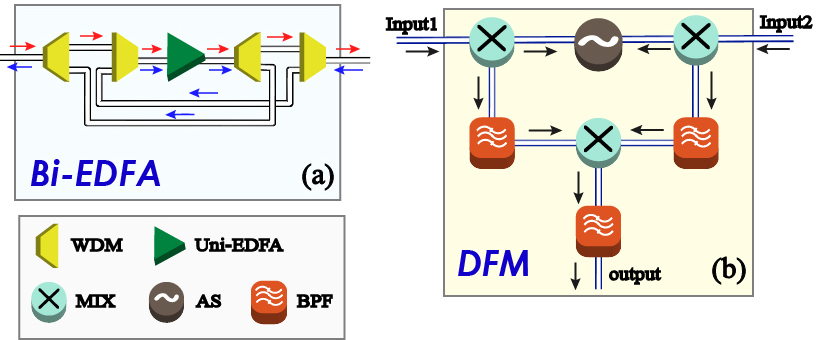}
\captionsetup{labelfont=bf}
\caption{Detailed configuration of the  (a) Bi-EDFA  and (b) DFM. Bi-EDFA: bidirectional erbium-doped fiber amplifier, DFM: dual frequency mixer, WDM: wavelength division multiplexer, AS: auxiliary source, MIX: frequency mixer, Uni-EDFA: unidirectional erbium-doped fiber amplifier, BPF: band-pass filter.}
\end{figure}

\subsection{Long-haul radio frequency transfer}
\begin{figure}[htbp]
\centering\includegraphics[width=10.5cm]{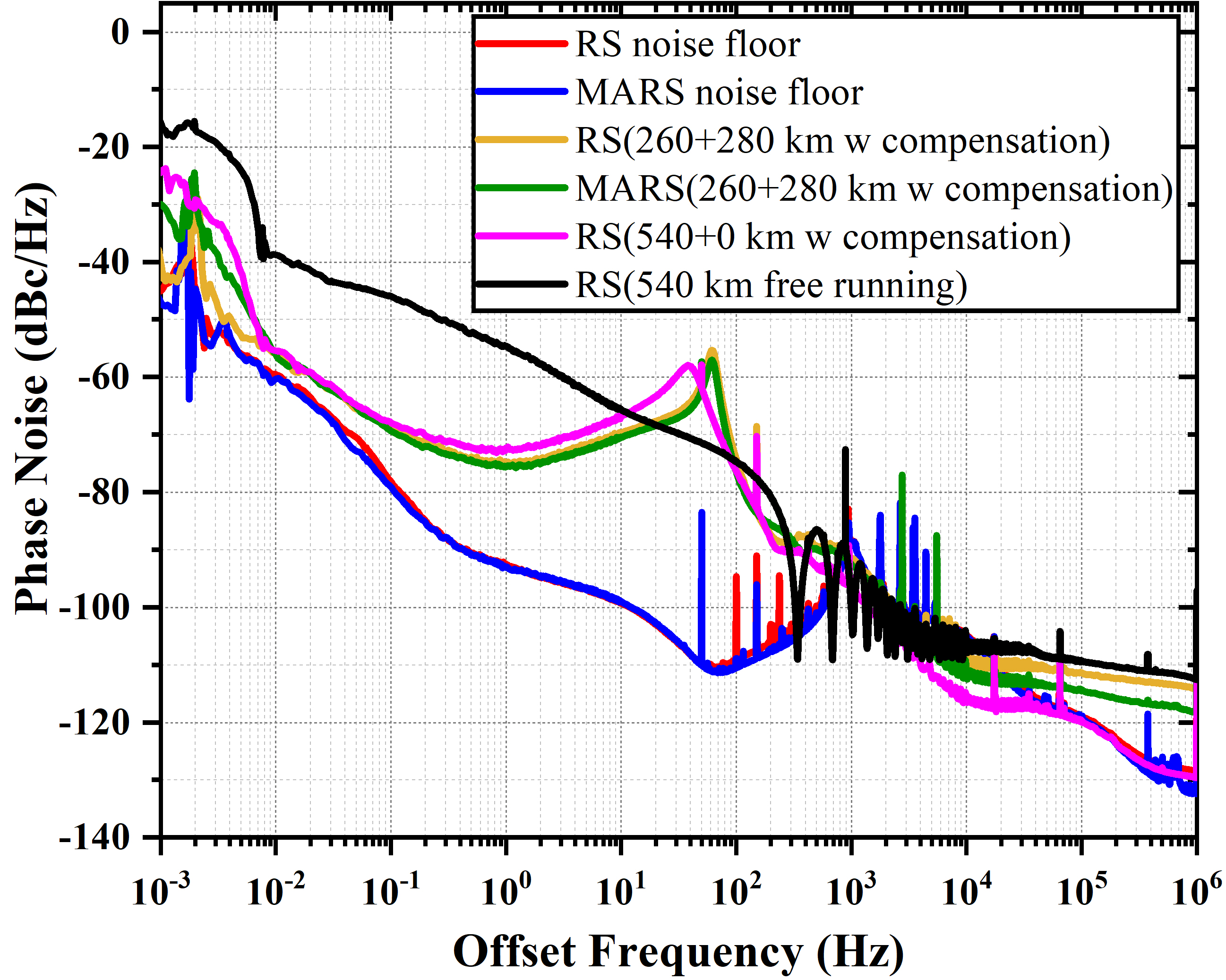}
\captionsetup{labelfont=bf}
\caption{Measured phase noise PSDs for the different system configurations: the system noise floor RS output (red curve), the system noise floor MARS output (blue curve), compensated 260+280 km RS output (yellow curve), MARS output with the 260+280 km fiber link (green curve), compensated 540+0 km RS output (pink curve), 540 km free running link (black curve). RS: remote site, MARS: multiple-access relay station.}
\end{figure}

Figure 3 illustrates the measured single-sideband (SSB)  phase noise power spectral densities (PSDs) of the system under different configurations. The RS and MARS back-to-back noise floors are measured by replacing $L_0$ and $L_1$ in the proposed scheme with short fiber patch cords. As a comparison, we also measure the long-haul RF transfer with the single-stage configuration as similar in Ref \cite{12,13}, by replacing $L_0$ and $L_1$ with 540 km SMFs and short fiber patch cord in the proposed scheme. The 540 km free running link is evaluated by transferring a 1 GHz signal to the RS without adding MARS and round-trip phase correction. As shown in Fig. 3, the SSB phase noise of the 540 km free running link is obviously higher than the compensated 260+280 km for RS output (about -68 dBc at 0.1 Hz, -74 dBc at 1 Hz) and MARS output (about -69 dBc at 0.1 Hz, -75 dBc at 1 Hz) at the offset frequency of less than 10 Hz, due to fiber delay fluctuations caused by the temperature variations. In comparison with the 540+0 km transfer for RS output, the SSB phase noise of the compensated 260+280 km for RS output is lower within 0.1-50 Hz, demonstrating that the proposed RF transfer scheme relieves the limitation of system's SNR over long-haul distribution. Moreover, the 540+0 km transfer for RS output has a smaller compensation bandwidth, which can be observed from the position of the bump in Fig. 3. In the range of 0.001-0.003 Hz, all the measurements have the sharp spikes, which are mainly determined by impact of the environmental temperature fluctuations on outside loop devices \cite{21}. The SSB phase noise of all compensation configurations has a bump observed at about 40, 60 or 1000 Hz. This part noise is mainly determined by the phase noise of the RF signal, the PLL parameters and the propagation delay from the fiber and so on \cite{20}. 

\begin{figure}[htbp!]
\centering\includegraphics[width=10.5cm]{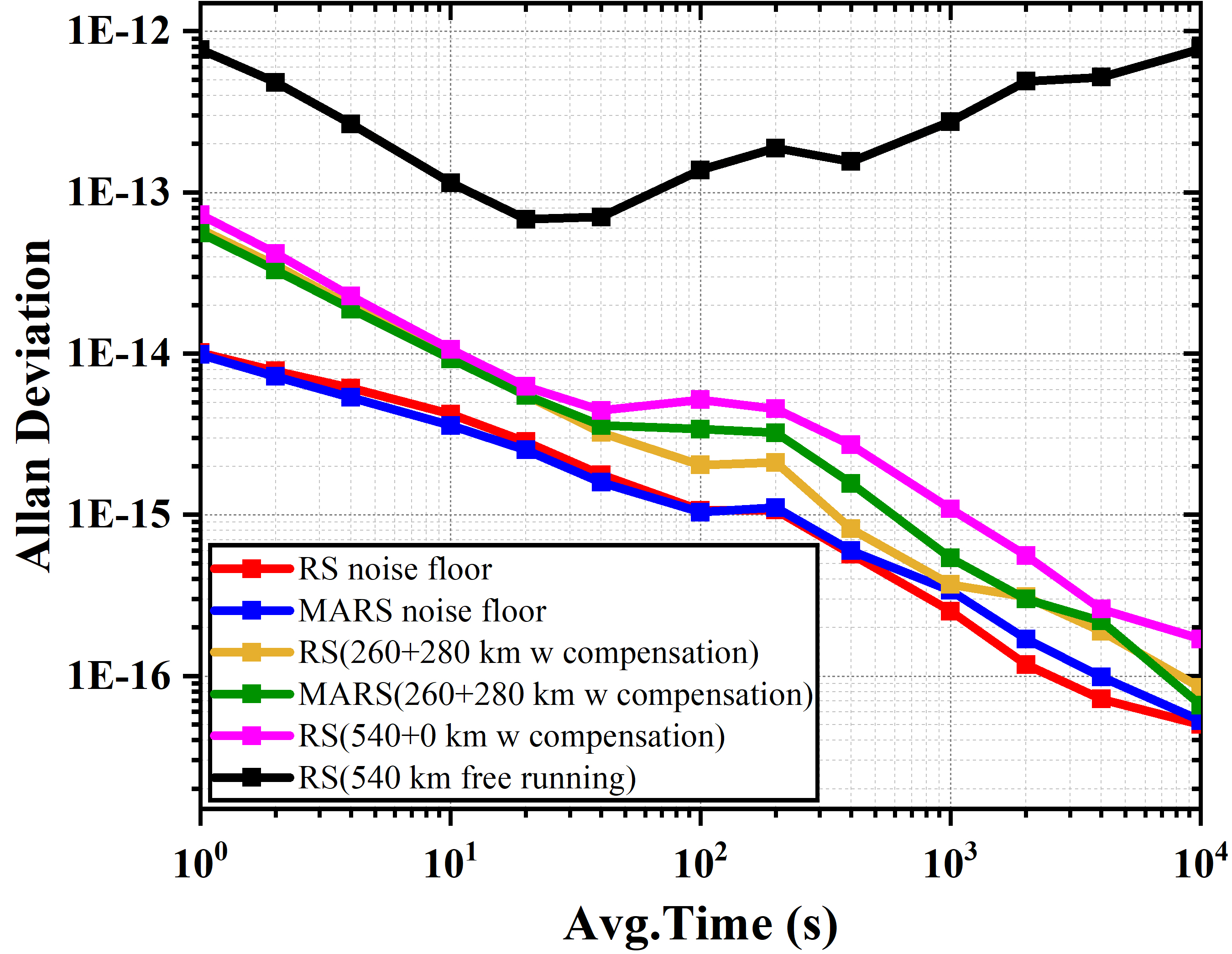}
\captionsetup{labelfont=bf}
\caption{Measured Allan deviations for the different transfer configurations: the system noise floor RS output (red curve), the system noise floor MARS output (blue curve), compensated 260+280 km RS output (yellow curve), MARS output with the 260+280 km fiber link (green curve), compensated 540+0 km RS output (pink curve), 540 km free running link (black curve). RS: remote site, MARS: multiple-access relay station.}
\end{figure}

The measured fractional frequency instability in terms of Allan deviations (ADEVs) for different system configurations are illustrated in Fig. 4. It can be observed that with the implementation of 260+280 km fiber noise compensation for the RS (MARS) output, the proposed RF transfer scheme achieves a fractional frequency instability of $5.9\times10^{-14}$ ($5.6\times10^{-14}$) at 1 s and reaches $8.5\times10^{-17}$ ($6.6\times10^{-17}$) at 10,000 s. In our experiment, we observe that the long-term stability of 260+280 km transfer for RS output is improved by approximately four orders of magnitude in contrast with 540 km free running link. Compared with the 540+0 km RF transfer for RS output, the ADEV of the compensated 260+280 km configuration for the RS output is improved from $7.3\times10^{-14}$ to $5.9\times10^{-14}$ at the averaging time of 1 s, indicating that the proposed scheme with the assistance of MARS can effectively improve the system’s SNR. The ADEV of the compensated 260+280 km for RS output is clearly higher than the system noise floor for the RS output ($1\times10^{-14}$ at 1 s, $5\times10^{-17}$ at 10,000 s), which is mainly attributed to the non-reciprocal noise in the fiber link and ASE noise introduced by the Bi-EDFAs \cite{13}. As the outside loop of the system is easily susceptible to the temperature fluctuations, the ADEVs of the different system configurations decrease slowly within 40-200 s, which correspond to the sharp spikes at the range of $0.001-0.003\ \rm{Hz}$ in Fig. 3.

\section{Discussion}
\subsection{The impact of MARS on the system in different positions}
Since the positions of stations in optical fiber communication network are usually not uniformly distributed \cite{OFT}, whether the MARS position will affect the transfer system needs to be carefully investigated. Here, we theoretically model the effect of the MARS position on the fractional frequency instability of the fiber-optic RF transfer system with two fiber sub-links. Based on the PLL theory and fiber-induced noise model \cite{23,24,25,26,Dong, hu2020passive}, the residual phase noise PSD of the compensation system can be obtained by extracting the phase of the beat note between the frequency standard and the remote signal, which can be denoted as,
\begin{equation}
\begin{aligned}
{S_{Residual}}(\omega ) =& {\left| {\frac{{{G_1}{e^{ - j\omega {\tau _1}}}\sqrt {2\left[ {1 - \rm{sinc}\left( {2\omega {\tau _0}} \right)} \right]} }}{{1 + {G_1}{e^{ - j2\omega {\tau _0}}} + {G_1}{e^{ - j2\omega {\tau _1}}}}}} \right|^2}{S_{p0}}(\omega )\\&
{\rm{+ }}{\left| {1 - \frac{{{G_1}{e^{ - j\omega {\tau _1}}}\sqrt {2\left[ {1 + \rm{sinc}\left( {2\omega {\tau _1}} \right)} \right]} }}{{1 + {G_1}{e^{ - j2\omega {\tau _0}}} + {G_1}{e^{ - j2\omega {\tau _1}}}}}} \right|^2}{S_{p1}}(\omega )\\&
{\rm{+ }}{\left| {1-\frac{{{2G_1}{e^{ - j\omega {\tau _0}}}{e^{ - j\omega {\tau _1}}}}}{{1 + {G_1}{e^{ - j2\omega {\tau _0}}} + {G_1}{e^{ - j2\omega {\tau _1}}}}}} \right|^2}{S_{RF}}(\omega ) + {S_{noise\_floor}}(\omega )\label{11}.
\end{aligned}
\end{equation}
where $\tau_0$ and $\tau_1$ are propagation delays of the front and rear fiber sub-links, $G_1=(K_P+K_I/{j\omega})\cdot K_{PFD}\cdot K_{VCO}/j\omega$ is the transfer function of PLL unit, $S_{p0} (\omega)$, $S_{p1} (\omega)$, $S_{RF} (\omega)$ and $S_{noise\_floor} (\omega)$ are the SSB phase noise induced by the front fiber sub-link, the rear fiber sub-link, frequency standard and back-to-back system. Furthermore, the SSB phase noise caused by the fiber link has been experimentally verified to be denoted in the form of a Power-laws spectrum, i.e., $S_{p} (\omega)={\omega_r}^{2}L\sum_{\alpha=-3}^{0}{h_{\alpha}(\omega/{2\pi})^\alpha}$, where $h_{\alpha}$ holds for the coefficient of the corresponding $\alpha$-order of noise \cite{23}. As the back-to-back system noise floor contains a part of the noise introduced by the RF signal, and this noise has little influence on the system. Therefore, we ignore the influence of this item in the simulation. 
\begin{figure}[htbp]
\centering\includegraphics[width=10.5cm]{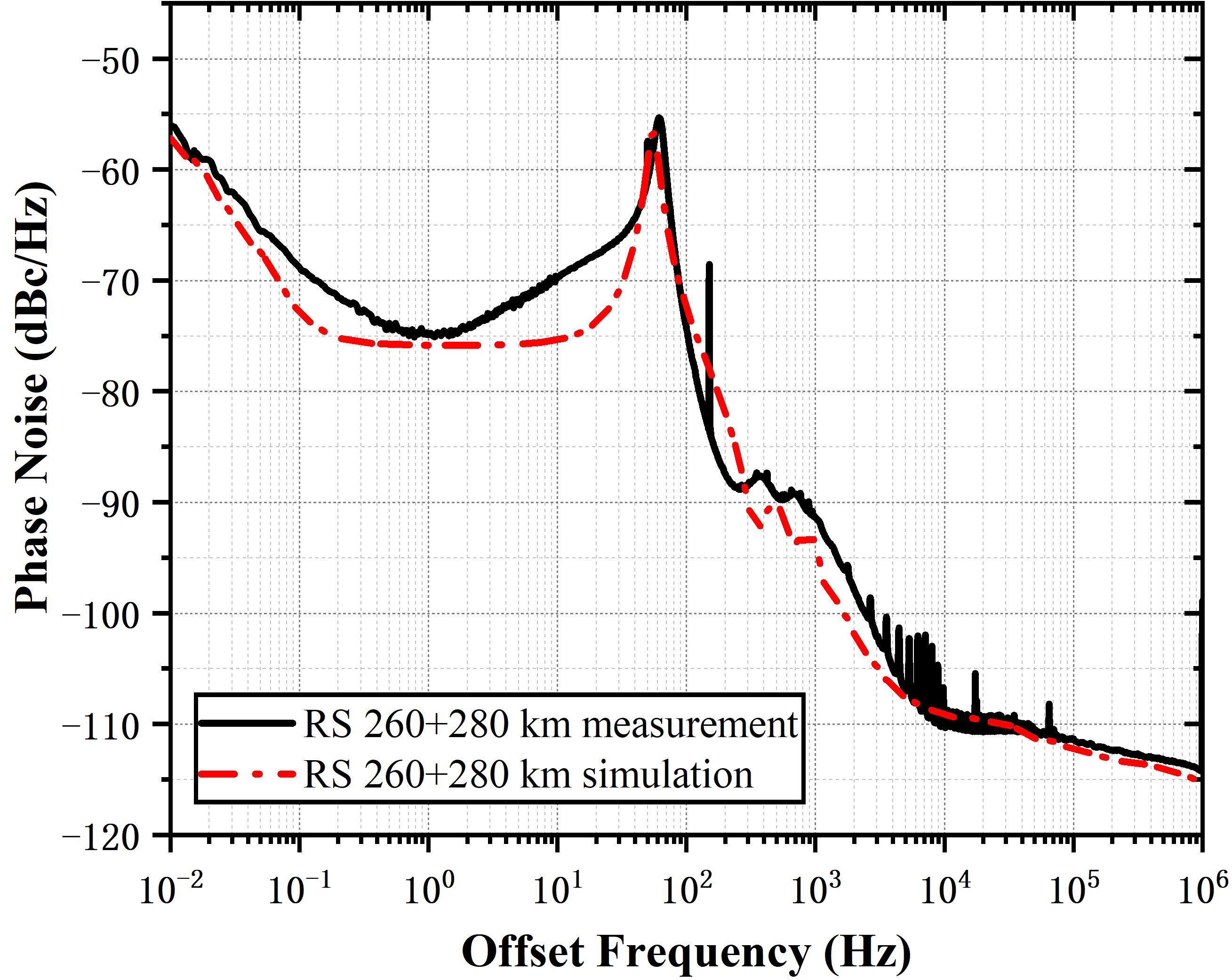}
\captionsetup{labelfont=bf}
\caption{Measured phase noise PSD for the compensated 260+280 km RS output (black curve) and simulated phase noise PSD for the compensated 260+280 km RS output (dashed red curve).}
\end{figure}

To verify the effectiveness of the simulation model, we compare the measured and simulated results for the compensated 260+280 km RS output with the help of the Eq. \ref{11}. As several Bi-EDFAs are adopted to boost the attenuation of the long fiber link \cite{12,13,22}, we add the ASE noise coming from the Bi-EDFAs to the system noise floor $S_{noise\_floor} (\omega)$ in the simulation model. According to the PLL theory, the ASE noise can be equivalently regarded as the additive white noise \cite{13,22} superimposed onto the phase detector in PLL unit. This part of the noise can be multiplied by the corresponding transfer function. Here we set the white noise introduced by the ASE to -78 dBc/Hz with the assistance of the model reported in Ref. \cite{22,EDFA}. As can be seen from Fig. 5 that the measured result is close to the simulated result at the high frequency range ( > 50 Hz) and the slight difference between the measured and simulated at low frequency range ( < 50 Hz) could be caused by the deviation of the ASE noise model, the non-reciprocal noise and the noise introduced by polarization mode dispersion and so on.

\begin{table}[!h]
\centering
\begin{scriptsize}
\captionsetup{labelfont=bf}
\caption{\label{Table2}\textbf{Relevant parameters used in the simulation of the compensation system }}
\begin{tabular}{ccc} \hline
Parameter & Symbol & Value \\ \hline
Proportion coefficient of loop filter in PLL unit & $K_P$ & $800$ \\
Integral coefficient of loop filter in PLL unit & $K_I$ & $25\times10^{4}\ \rm{rad/s}$ \\
Gain coefficient of phase frequency detector in PLL unit & $K_{PFD}$ & $6\times10^{-2}\ \rm{V/rad}$ \\
Tuning sensitivity of VCO in PLL unit & $K_{VCO}$ & $32\ \rm{rad/(s\cdot V)}$ \\
Propagation velocity in optical fiber & $c_n$ & $2\times10^8 \ \rm{m/s}$ \\
Corresponding -3-order coefficient to the fiber-induced noise & $h_{-3}$ & $1\times10^{-34}\ \rm{km^{-1}}$ \\
Corresponding -2-order coefficient to the fiber-induced noise & $h_{-2}$ & $2\times10^{-34}\ \rm{km^{-1}Hz^{-1}}$ \\
Corresponding -1-order coefficient to the fiber-induced noise & $h_{-1}$ & $6\times10^{-33}\ \rm{km^{-1}Hz^{-2}}$\\
Corresponding 0-order coefficient to the fiber-induced noise & $h_{0}$ & $5\times10^{-37}\ \rm{km^{-1}Hz^{-3}}$ \\\hline
\end{tabular}
\end{scriptsize}
\end{table}

\begin{figure}[!ht]
\centering\includegraphics[width=13cm]{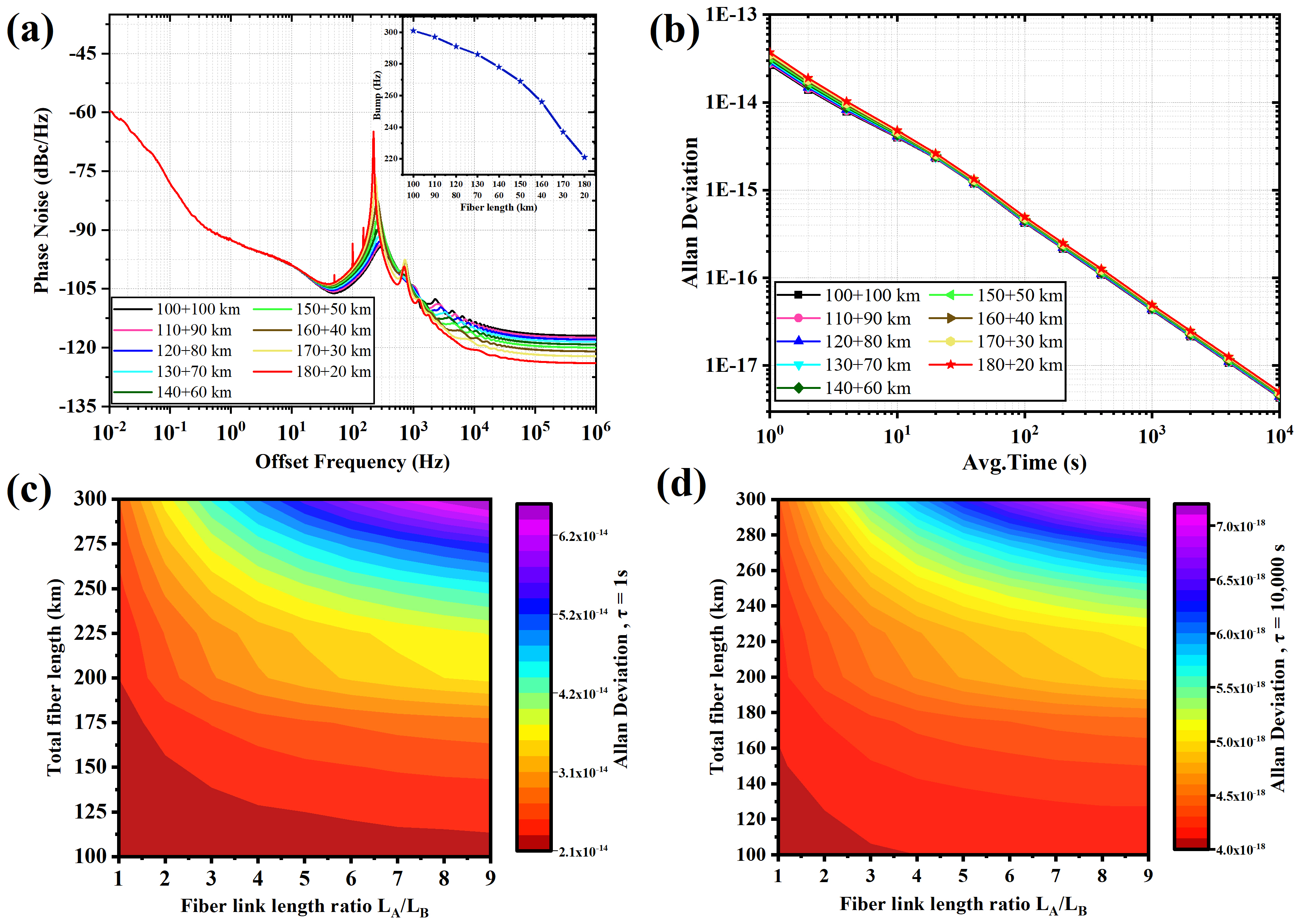}
\captionsetup{labelfont=bf}
\caption{(a) The simulated phase noise PSDs of the RF transfer systems over different link configurations with a total length of 200 km. The inset shows the position of the bumps in different fiber link configurations. (b) The simulated Allan deviations of the RF transfer systems over different link configurations with a total length of 200 km. (c) Total fiber length vs fiber link length ratio for different ADEVs at the averaging time of 1 s. (d) Total fiber length vs fiber link length ratio for different Allan deviations at the averaging time of 10,000 s.}
\end{figure}

Based on the aftermentioned model, we evaluate the impact of MARS location on the RF transfer system. The relevant parameters used for the calculation are listed in Table 2. It can be seen from Fig. 6(a) that, the fiber-induced phase noise at low frequency band is effectively suppressed by the compensation system, so the phase noise PSDs of different link configurations within 10 Hz are limited by the system noise floor. In the range of 10-100 Hz, as the length of a certain fiber sub-link increases, the compensation system gradually deteriorates the fiber-induced noise suppression capability at the mid-band frequency due to the decrease of the compensation bandwidth (the position of the bump is shifted towards low frequency range). At the high frequency range ( >$10^3$ Hz), with the distance of a certain fiber sub-link increases, the compensation system has more significant effect on high frequency link noise suppression. The trends of these simulated results are consistent with the experimental measured results in Fig. 3. According to the conversion relationship between phase noise PSD and ADEV \cite{27}, the ADEV results under different link configurations can be obtained, as shown in Fig. 6(b). The short-term stabilities of the 100+100 km and 120+80 km fiber links are $2.7\times10^{-14}$ and $2.9\times10^{-14}$ at the averaging time of 1 s, indicating that the MARS position has little effect on the system's performance when the ratio of the front and rear fiber sub-links is around $1:1$. The difference in long-term stabilities of the system (at the integration time of 10,000 s) become smaller under different fiber link configurations, due to the fact that the system noise floor is dominant in the low frequency range. Additionally, it is observed from Fig. 6(c) and Fig. 6(d) that with the increase of the total fiber length, the influence of the link length ratio (the position of the MARS) on the system stability also gradually increases. When the total distribution distance is 300 km, the ADEV at the averaging time of 1 s (10,000 s) deteriorates from $2.7\times10^{-14}$ ($4.4\times10^{-18}$) to $6.6\times10^{-14}$ ($7.3\times10^{-18}$) for the 1:1 link length configuration compared to the 9:1 link configuration. Therefore, it is obviously found that when the total transfer distance is less than 250 km, the MARS at different positions has little effect on the system, that is, the sensitivity of the MARS to position is not strong. However, when faced with the application requirements of ultra-long-haul RF distribution, the MARS can be placed close to the middle of the front and rear fiber sub-links.

\subsection	{Ultra-long-haul RF transfer by adopting N MARSs}

\begin{figure}[htbp]
\centering\includegraphics[width=10.5cm]{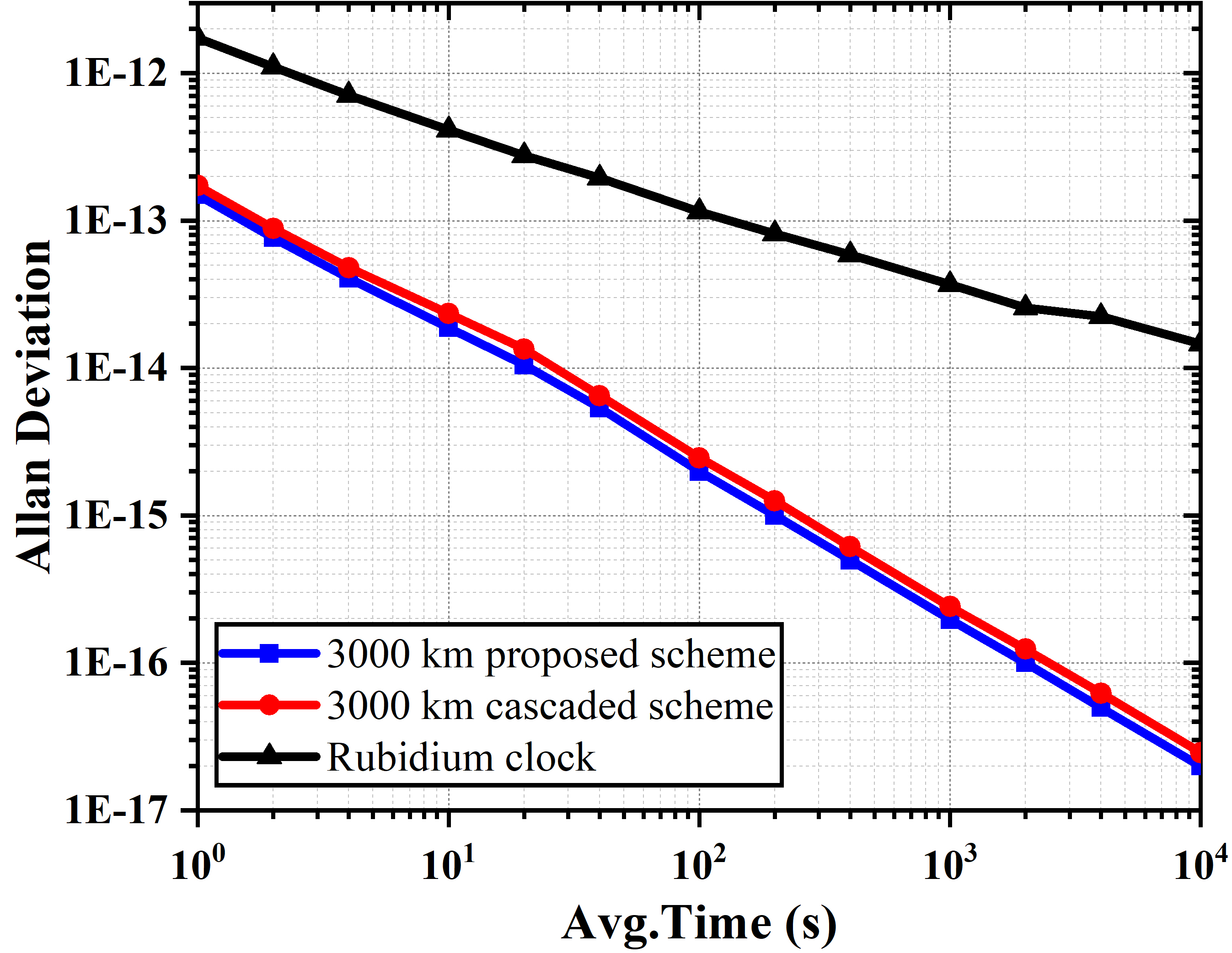}
\captionsetup{labelfont=bf}
\caption{Simulated Allan deviation results of the proposed system with 29 MARSs over 3000 km distribution link (blue curve) and the cascaded scheme with 30 stages over 3000 km distribution link (red curve). MARS: multiple-access relay station.}
\end{figure}

For ultra-long-haul RF transfer via fiber-optic link, it is indispensable to introduce more MARSs. Here, we theoretically analyze the potential of this proposed scheme with the assistance of $N$ MARSs by modeling the residual phase noise PSD of the RF transfer system. The idea of modeling is to bring the phase relationship obtained by the first $N-1$ MARSs into the $N$th MARS through $N-1$ iterations, that is, substituting Eq. \ref{12}-\ref{14} into Eq. \ref{15}. The subscript $k$ in Eq. \ref{13} represents arbitrary MARS except for the first and last MARSs. $S_{noise\_floor\_N}$ stands for the back-to-back noise floor of the system with $N$ MARSs. Based on the residual noise model of $N$ MARSs, we simulate the ultra-long-haul RF transfer via 3000 km fiber-optic link. Assuming that there are 29 MARSs placed at equal distances on the transmission link, that is, each short fiber sub-link is 100 km. At the same, we keep the PLL parameters of each MARS consistent. It can be observed from Fig. 7 that the proposed scheme with 29 MARSs is expected to achieve short-term stability less than $1.5\times10^{-13}$ at the integration time of 1 s and long-term stability less than $1.9\times10^{-17}$ at the integration time of 10,000 s via 3000 km fiber-optic link, which is far better than that of rubidium clocks \cite{12}. As a comparison, we evaluate performance of the conventional 30-stages cascaded scheme over 3000 km fiber link with the help of phase noise model reported in Ref. \cite{Dong}. It can be seen from the simulated results that the proposed scheme can obtain slight better stability than that of the conventional cascaded scheme in the 3000 km link. More importantly, our proposed technique can largely simplify the configuration at each repeater station.

\begin{equation}
{\varphi _{c1}}\left( \omega \right) = \frac{{ - {G_1}}}{{1 + {G_1}{e^{ - j2\omega {\tau _0}}}}}\left[ \begin{array}{l}
{\varphi _{c2}}\left( \omega \right){e^{ - j\omega {\tau _1}}} + {\varphi _{p1}}\left( \omega \right) - {\varphi _{RF}}\left( \omega \right){e^{ - j\omega {\tau _0}}}\\
 - \sqrt {2\left[ {1 - \rm{sinc}\left( {2\omega {\tau _0}} \right)} \right]} {\varphi _{p0}}\left( \omega \right)
\end{array} \right]\label{12}.
\end{equation}

\begin{equation}
{\varphi _{ck}}\left( \omega \right) = {-G_k}\left[ \begin{array}{l}
{\varphi _{ck + 1}}\left( \omega \right){e^{ - j\omega {\tau _k}}} + {\varphi _{pk}}\left( \omega \right)\\
 - {\varphi _{ck - 1}}\left( \omega \right){e^{ - j\omega {\tau _{k - 1}}}} - {\varphi _{pk - 1}}\left( \omega \right)
\end{array} \right]\label{13}.
\end{equation}

\begin{equation}
{\varphi _{cN}}\left( \omega \right) = \frac{{ - {G_N}}}{{1{\rm{ + }}{G_N}{e^{ - j2\omega {\tau_{N}}}}}}\left[ \begin{array}{l}
\sqrt {2\left[ {1 + {\rm{sinc}}\left( {2\omega {\tau_{N}}} \right)} \right]} {\varphi _{pN}}\left( \omega \right)\\
 - {\varphi _{cN - 1}}\left( \omega \right){e^{ - j\omega {\tau _{N - 1}}}} - {\varphi _{pN-1}}\left( \omega \right)
\end{array} \right]\label{14}.
\end{equation}

\begin{equation}
{S_{Residual}}(\omega ) = \left\langle {{{\left| {{\varphi _{cN}}\left( \omega \right){e^{ - j\omega {\tau _N}}} + {\varphi _{pN}}\left( \omega \right)} \right|}^2}} \right\rangle + {S_{noise\_floor\_N}}(\omega )\label{15}.
\end{equation}

\section{Conclusions}
In conclusion, we present a long-haul RF transfer scheme by adopting MARSs. The proposed scheme has the capability to independently compensate the fiber-induced noise of each fiber sub-link, which effectively solves the limitation of compensation bandwidth for long-haul transfer. This MARS shares the same modulated optical signal for the front and rear fiber sub-links, which can not only simplify the configuration of RF transfer configuration but also provide the multiple-access capability. Experimentally, we demonstrate a 1 GHz signal transfer based on the proposed scheme by adopting one MARS via 260+280 km SMFs. The stable 1 GHz signal transferred to the RS has the fractional frequency instability of less than $5.9\times10^{-14}$ at the integration time of 1 s and $8.5\times10^{-17}$ at 10,000 s. At the MARS, the stable 1 GHz signal with the stabilities of less than $5.6\times10^{-14}$ and $6.6\times10^{-17}$ at 1 s and 10,000 s can also be obtained. Moreover, we for the first time theoretically model the effect of the MARS position on the fractional frequency instability of the fiber-optic RF transfer system with two fiber sub-links, demonstrating that the MARS position has little effect on system's performance when the ratio of the front and rear fiber sub-links is around $1:1$. Promisingly, the proposed scheme with great scalability are potential for ultra-long-haul RF transfer by arbitrarily adding the same MARSs on the main fiber link.

\begin{backmatter}
\bmsection{Funding}
This work was supported by the National Natural Science Foundation of China (NSFC) (62120106010, 61905143), the Zhejiang provincial Key Research and Development Program of China (2022C01156) and the National Science Foundation of Shanghai (22ZR1430200).

\bmsection{Disclosures}
The authors declare no conflicts of interest.

\bmsection{Data availability}
Data underlying the results presented in this paper are not publicly available at this time but may be obtained from the authors upon reasonable request.

\end{backmatter}







\end{document}